\documentclass{article}
\usepackage{latexsym}
\usepackage{amsmath}
\usepackage[dvips]{graphicx}
\newcommand{\beq}{\begin{equation}}
\newcommand{\eeq}{\end{equation}}
\newcommand{\beqa}{\begin{eqnarray}}
\newcommand{\eeqa}{\end{eqnarray}}
\newcommand{\bseq}{\begin{subequations}}
\newcommand{\eseq}{\end{subequations}}

\newcommand{\trm}{\textrm}
\title{Pseudoscalar Meson Temporal Correlation\\ Function \\
 for Finite Momenta in HTL approach\footnote{Talk given at the MESON2006,
 9thInternational Workshop on Meson Production, Properties and Interaction,
Krak\'ow, Poland, 
9-13 June 2006}}
\author{\\{Piotr Czerski}\\
\\
{\it Institute of Nuclear Physics} \\
{\it Polish
Academy of Sciences}\\
{\it ul. Radzikowskiego 152 }\\
{\it PL-31-342 Krak\'ow, Poland}}
\begin{document}
\maketitle
\begin{abstract}
The temporal pseudoscalar meson correlation function in a QCD plasma
 is investigated in a range of temperatures exceeding $T_c$ and first
 time for a finite momenta which is of the experimental interest.
 The imaginary time formalism is employed for the finite temperature
  calculations. The behavior of the meson spectral function and
   of the temporal correlator is studied in the HTL approximation,
    where one replaces the free thermal quark propagators with
     the HTL resumed ones.\\ \\
     {\it Key words}:Finite temperature QCD, Quark Gluon Plasma, 
 Meson correlation function, 
Meson spectral function, Finite momentum, HTL approximation.
  \\ 
 {\it PACS}: 10.10.Wx, 11.55.Hx, 12.38.Mh, 14.65.Bt, 14.70.Dj, 25.75.Nq    
\end{abstract}
\section{Introduction}	
Present contribution is based on the article \cite{my} done in collaboration
with W.M. Alberico, A. Beraudo and A. Molinari. We compute the Meson Spectral
 Function (MSF) in the HTL approximation. The degrees of freedom we 
deal with are light quarks (massless) and gluons 
and our results refer to the case of zero chemical potential, a condition
 which is expected to be realized in the heavy ion
 experiments at RHIC and even better in the future experiments at LHC.
  The calculations are performed in the imaginary time formalism, taking at
the end a proper analytical continuation to real frequencies when required.
The zero momentum MSF \cite{berry} at small values of the energy 
displays \emph{Van Hove singularities}, which exist because of the presence
 of the plasmino mode in the HTL quark propagator. 
  Divergences appear in the density of states giving rise to peaks in the MSF.
   The possible experimental relevance of such singularities was analyzed in
 Ref.~\cite{bra}, where the back-to-back dilepton production rate was 
 evaluated
 in terms of the zero momentum MSF in the vector channel.
The full information on the momentum and temperature behavior of the
 mesonic excitations in the QGP phase is encoded in their spectral function
 for any energy and momenta.
\section{Finite temperature meson spectral function}
The analytical expression for MSF is:
\begin{multline}
\sigma^{\trm{ps}}_{HTL}(\omega,\mathbf{p})\!=\!2N_c\int\!\frac{d^3k}{(2\pi)^3}
(e^{\beta\omega}-1)
\int\limits_{-\infty}^{+\infty}d\omega_1\int\limits_{-\infty}^{+\infty}
d\omega_2\tilde{n}(\omega_1)\tilde{n}
(\omega_2)\delta(\omega-\!\omega_1-\!\omega_2)\times\\
\times\left\{(1+\mathbf{\hat{k}\cdot\hat{q}})[\rho_+(\omega_1,k)\rho_+(\omega_2,q)
+\rho_-(\omega_1,k)
\rho_-(\omega_2,q)]+\right.\\
+\left.(1-\mathbf{\hat{k}\cdot\hat{q}})[\rho_+(\omega_1,k)\rho_-(\omega_2,q)
+\rho_-(\omega_1,k)
\rho_+(\omega_2,q)]\right\} ,\label{eq:sigmapcompl}
\end{multline}
where $\beta=1/T$ and $\tilde{n}(\omega)$ is a Fermi distribution.
The HTL quark spectral function
 $\rho_{\pm}(\omega,k)$, whose expression in given in Ref.~\cite{my},
  reflects the singularities of the quark propagator in the
   complex $\omega$-plane. These lie on the real $\omega$-axis. At a given
    value of the spatial momentum $k$, in the time-like domain
     ($\omega^2>k^2$) discrete poles are associated to quasiparticle 
     excitations; a cut for space-like momenta ($\omega^2<k^2$) is related to
      the Landau damping.

The thermal meson propagator can be expressed through the
 spectral function $\sigma_{HTL}$: 
\beq
G_{HTL}(\tau,\mathbf{p})=-\frac{1}{\beta}\sum_{n=-\infty}^{+\infty}e^{-i\omega_n\tau}
\int\limits_{-\infty}^{+\infty}
d\omega\frac{\sigma_{HTL}(\omega,\mathbf{p})}{i\omega_n-\omega}=\int\limits_0^{+\infty}d\omega~
\sigma_{HTL}(\omega,\mathbf{p})K(\omega,\tau)
\label{eq:gtau}
\eeq
with $\tau\in[0,\beta]$, the kernel 
$K(\omega,\tau)=\frac{\cosh[\omega(\tau-\beta/2)]}{\sinh(\omega\beta/2)}$ and
the sum over the Matsubara frequencies with a standard contour
 integration in
the complex $\omega$ plane \cite{lb,bi} is performed \cite{mus}.

We shall denote with $q_+$ ($\bar{q}_+$) the normal quark 
(antiquark) mode, with $q_-$ ($\bar{q}_-$) the plasmino (antiplasmino)
 mode and with $M$ the excitation carrying the quantum numbers of a 
 pseudoscalar meson.
\section{Results}
\begin{figure}[!htp]
\begin{center}
\includegraphics[clip,width=0.7\textwidth]{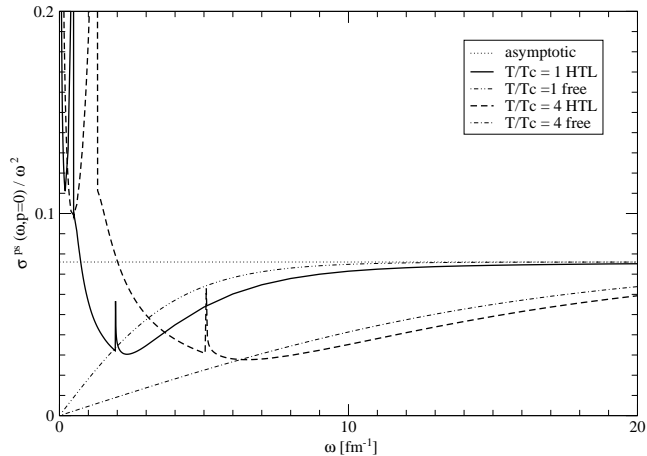}
\caption{The zero momentum HTL and free MSF (divided by $\omega^2$) as a 
function of $\omega$ for two different temperatures.
 The asymptotic high energy plateau is also shown.}
\label{Fig:spec_p0} 
\end{center}
\end{figure}
In Fig. \ref{Fig:spec_p0} we plot the spectral function 
divided 
by $\omega^2$. Two different temperatures are considered. One can appreciate the 
 dramatic difference in the behavior at low energy between the free curves 
 (which vanish at $\omega=0$) and the HTL ones (which diverge for $\omega\to0$).
  One can also recognize the Van Hove singularities in the HTL curves
   arising from a divergence in the density of states due to the minimum in 
   the plasmino dispersion relation. All the curves approach for large values 
   of $\omega$ the same high energy, temperature independent, plateau.
We now investigate how things change at finite spatial momentum. 
\begin{figure}[!htp]
\begin{center}
\includegraphics[clip,width=0.7\textwidth]{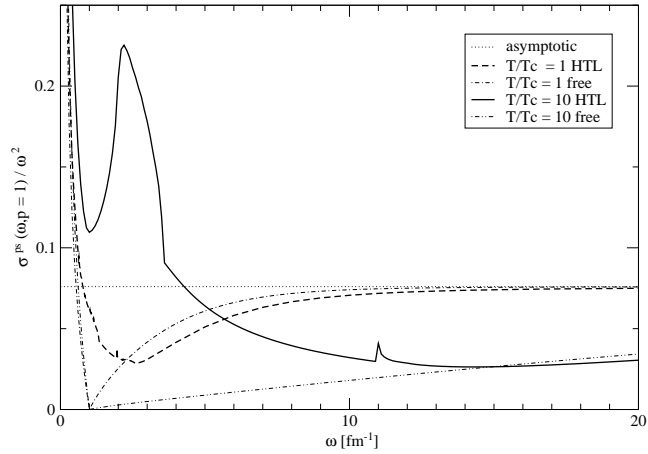}
\caption{The $p=1$ fm$^{-1}$ HTL and free MSF (divided by $\omega^2$)
 as a function of $\omega$ for two different temperatures. 
 The asymptotic high energy plateau is also shown.}
\label{Fig:spec_p1} 
\end{center}
\end{figure}
In Fig. \ref{Fig:spec_p1} we plot the 
pseudoscalar MSF for $p=1$fm$^{-1}$ at different 
temperatures. The non-interacting result vanishes on the 
light-cone, at variance with the HTL curves, which stays finite there. 
 On the other hand when $\mathbf{p}$ is finite both the free and the interacting 
 curves diverge as $\omega\to0$. Furthermore in the HTL case the Van Hove 
 singularities are washed-out by the angular integration. 
 Finally both the free and the HTL finite momentum MSF 
 approach the asymptotic plateau for large values of $\omega$. 
\begin{figure}[!htp]
\begin{center}
\includegraphics[clip,width=0.70\textwidth]{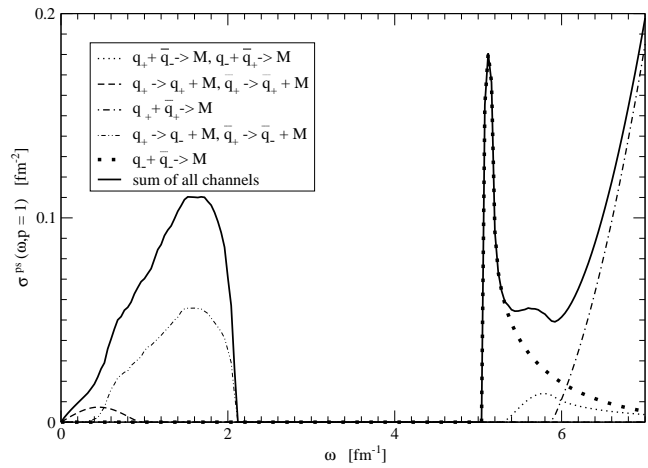}
\caption{The major processes contributing to the pole-pole term
 of the HTL MSF. The plot is given for the case $p=1$ fm$^{-1}$ and $T=2T_c$.}
\label{Fig:zoom}
\end{center}
\end{figure}

We can also see Fig. \ref{Fig:zoom} how the different processes 
contribute to the pole-pole term \cite{my}, at a given value of the 
temperature and of the spatial momentum, to the HTL MSF. 
In analogy to the zero-momentum case studied in Ref. \cite{berry,mus}, it turns 
out that the dominant process at low energy is the decay 
$q_+\rightarrow q_-+M$; then there is a wide gap for an intermediate range
 of energy till when the plasmino-antiplasmino annihilation starts contributing.
  Such a process initially gives a quite large contribution due to the large 
  density of states; then it decreases rapidly with the energy because of the 
  very small value of the plasmino residue.
It appears that, for large enough frequencies, 
the dominant role is played by the quark-antiquark annihilation. Such a 
process starts contributing for $\omega$ larger than a threshold depending 
on the thermal gap mass $m_q$ acquired by the quarks in the thermal bath.
On the other hand the processes $\bar{q}_- \rightarrow  
\bar{q}_- + M$ and $q_-\rightarrow q_+ + M$ 
 turn out to be totally negligible, due to the low value of 
the plasmino residue and to the very small available phase space.

\begin{figure}[!htp]
\begin{center}
\includegraphics[clip,width=0.7\textwidth,angle=270]{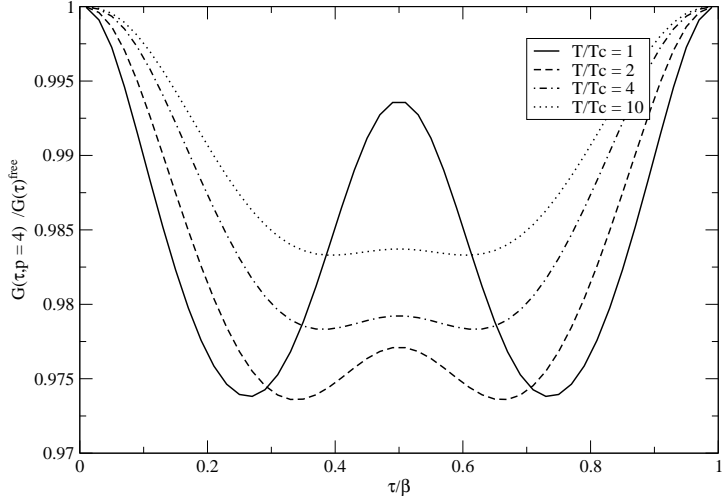}
\caption{The ratio $G_\trm{HTL}(\tau)/G_\trm{free}(\tau)$ for different
 temperatures at $p=4$fm$^{-1}$.}
\label{Fig:htl_free4} 
\end{center}
\end{figure}

The finite momentum temporal correlator is defined in Eq. (\ref{eq:gtau}).
In order to assess the impact of the interaction it is convenient to consider
 the ratio $G_\trm{HTL}(\tau)/G_\trm{free}(\tau)$ between HTL and a free cases.
Such a ratio turns out to be finite for every value of $\tau$. 
The above ratio approaches 1 for $\tau\to 0$ 
   (or $\beta$). In fact, due to the structure of the thermal kernel given in
    Eq. \ref{eq:gtau}, in this limit the correlator $G(\tau)$ is dominated 
    by the high energy behavior of the MSF.   
\begin{figure}[!htp]
\begin{center}
\includegraphics[clip,width=0.7\textwidth,angle=270]{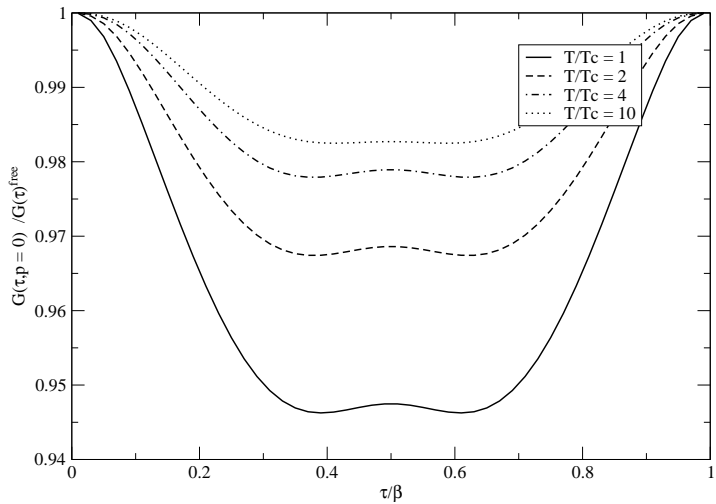}
\caption{The ratio $G_\trm{HTL}(\tau)/G_\trm{free}(\tau)$ for different
 temperatures at $p=0$fm$^{-1}$.}
\label{Fig:htl_free0} 
\end{center}
\end{figure}
We start by considering the zero momentum case, left part of
 Fig. \ref{Fig:htl_free4},
 showing the ratio for a range of temperatures from 
$T=T_c$ to $T=10T_c$.
Then we move to the finite momentum case, $p=4$fm$^{-1}$, right part of 
Fig. \ref{Fig:htl_free0}. As the temperature increases the ratio 
 moves closer to $1$, 
 reflecting the running of the coupling, and the bump at $\tau=\beta/2$ is 
 smeared out.
\section{Conclusions}
We have examined the impact of a finite value of the spatial momentum on the 
spectral density and on the temporal correlation function of a pseudoscalar 
meson for temperatures above the deconfinement phase transition and zero 
chemical potential. This amounts to study the properties of an excitation 
carrying the quantum number of a meson and propagating
 in the heat-bath frame.
In our treatment we employed HTL resummed quark propagators. This has allowed
 us to perform many calculations analytically \cite{my} and
  to identify the different physical processes contributing to the MSFs and
   to the temporal correlators, a task impossible to achieve, of course, 
   when the same quantities are evaluated on the lattice.
   
At zero 
momentum, the main features of the MSF are the presence of the Van Hove 
singularities and the 
radically different low energy behavior of the HTL predictions with respect 
to the free case. Such a contrast results particularly visible in the plot 
of $\sigma(\omega,0)/\omega^2$, where, for $\omega\to 0$ the HTL 
curve diverges while the free result vanishes.
At finite spatial momentum it turns out that the Van Hove singularities 
so prominent in the zero momentum case are smoothed out by the angular 
integration.
Another finding worth to be pointed out is that while the free MSFs vanish 
on the light-cone, the HTL curves stay finite.
The impact of a finite value of the spatial momentum has been 
investigated also for the HTL temporal correlator $G(\tau,p)$, for 
which we plotted the ratios with respect to the free 
results.
We 
found that the ratio of the HTL result with respect to the non-interacting 
correlator differs from one just for a few percent.

We hope our work can provide a complementary and independent approach to 
the studies of MSFs and temporal correlators performed on the lattice.

\end{document}